\documentclass[conference]{IEEEtran}
\IEEEoverridecommandlockouts

\usepackage{cite}
\usepackage{amsmath,amssymb,amsfonts}
\usepackage{graphicx}
\usepackage{textcomp}
\usepackage{xcolor}
\usepackage{booktabs}
\usepackage{url}
\usepackage{microtype}
\usepackage{listings}
\usepackage{tikz}
\usetikzlibrary{shapes.geometric, arrows.meta, positioning, fit}

\begin{document}

\title{A Graph-Native Bitemporal Memory Store for\\Conversational AI Agents}

\author{
  \IEEEauthorblockN{Alp Niksarli}
  \IEEEauthorblockA{Davidson College\\alniksarli@davidson.edu}
  \and
  \IEEEauthorblockN{Gopesh Baheti}
  \IEEEauthorblockA{Davidson College\\gobaheti@davidson.edu}
}

\maketitle

\begin{abstract}
Conversational AI agents commonly lack persistent memory across sessions. The obvious fixes like injecting full chat histories into the context window, or delegating to a third-party memory service, either exhaust the model's context budget or send personal data through infrastructure the user does not control. We describe a memory store that avoids both problems: an agent-local Neo4j property graph augmented with HNSW vector indexes and a full bitemporal data model. Each memory is stored as an immutable identity node linked to versioned content nodes carrying two closed-open time intervals—\emph{valid time} (when the fact was true in the world) and \emph{transaction time} (when the database recorded it). This design supports point-in-time semantic retrieval without physically overwriting history. Semantic edges between related memories are maintained automatically at write time using cosine similarity over 1024-dimensional embeddings. We evaluate the system on LongMemEval, a 500-question benchmark spanning six question types designed to stress long-term memory. Across 60 sampled questions, the current-state semantic search path achieves 46.7\% R@10 overall, rising to 80\% on knowledge-update questions. The time-travel path yields 80\% R@10 on knowledge-update but \emph{decreases} recall on temporal-reasoning questions (50\%\,→\,37.5\%), a consequence of post-filter dilution that points directly to a concrete design improvement. We discuss what these results reveal about the limits of pure retrieval for different question types and what each failure mode suggests for future work.
\end{abstract}

\section{Introduction}

Most deployed conversational agents operate without durable memory. Each session starts cold, and the agent has no access to what the user said a week ago unless the application developer explicitly provides it. The naive solution of prepending the entire conversation history to every prompt, works only at small scale. Token costs grow linearly with history length, and the model's ability to attend to relevant context degrades when the window is crowded with irrelevant turns~\cite{lostinthemiddle}.

Third-party AI memory services (Mem0, Zep, LangChain's \texttt{ConversationSummaryMemory}) address this by maintaining an external retrieval index. The agent queries it each turn and injects only the retrieved snippets into the context window. This is sensible engineering, but it moves a record of everything the user has said to a service the user does not control. For agents that handle health notes, financial information, or private correspondence, the privacy cost is real.

A local-first design sidesteps this. If the memory store is the agent's own database, i.e. running on the same machine or in a managed instance owned by the developer, retrieval latency is low and no personal data leaves the user's sphere. The architecture we describe is designed around this principle: Neo4j with Bolt over localhost is the default deployment target, and the implementation works against Neo4j Aura (cloud-managed) only because that is what we had available for testing.

Beyond the deployment question, memory systems for conversational agents face several harder problems. Standard vector stores assume static embeddings: an update overwrites the prior vector and the old state is gone. For many personal agent use cases this is wrong. If a user told the agent they take no medication in January and then mentioned a new prescription in March, both facts are historically interesting: what was true then, and what is true now, are different questions. A plain key-value or mutable vector store cannot distinguish them.

We address this limitation with a bitemporal schema built on top of Neo4j’s native HNSW vector indexing infrastructure. Our contributions are twofold: (1) an identity/version schema that enables temporal queries in Cypher while preserving efficient vector retrieval, and (2) a dual-index design that supports both low-latency access to the current state and retrieval over complete historical records without duplicating data. We evaluate the system using LongMemEval~\cite{longmemeval} rather than a synthetic benchmark, which provides a more realistic view of both the strengths and limitations of retrieval-based memory systems.

\section{Related Work}

\subsection{Agent Memory Paradigms}
Prior work generally divides LLM agent memory into three categories: parametric memory stored in model weights, in-context memory maintained within the prompt window, and retrieval-augmented memory that pulls information from external storage~\cite{rag}. In practice, common in-context methods—such as LangChain’s \texttt{ConversationBufferMemory} or direct system-prompt injection—work well only up to a certain scale, after which larger context windows begin to introduce higher latency and weaker coherence. Summarization-based approaches~\cite{memgpt} help reduce context length by compressing conversation history, but this process inevitably removes details that cannot later be recovered. MemGPT~\cite{memgpt} addressed this limitation with a paging-style architecture that shifts information between an in-context “main memory” and external storage. However, its retrieval mechanism still relies on a flat vector index without explicit temporal organization. Our approach is complementary to MemGPT: the paging framework it proposes could operate on top of the temporally structured storage system we introduce.

Additionally, third-party systems such as Mem0 and Zep provide managed memory APIs with automated extraction and retrieval capabilities. While these services are practical and technically capable, they require user conversations to be transmitted to external servers. For privacy-sensitive applications where conversational data must remain within a controlled environment, this requirement makes such approaches unsuitable.

\subsection{Vector Databases}
Pinecone, Weaviate, Chroma, and Qdrant all support approximate nearest-neighbor retrieval over dense embeddings using variants of HNSW~\cite{hnsw}. Despite their differences, these systems generally assume that each document corresponds to a single current embedding that reflects its latest/current state. Some platforms include limited lifecycle support—such as soft deletes in Weaviate or basic version tracking in Qdrant—but none provide a fully bitemporal query model capable of answering questions such as: “Which version of this fact was valid at time $t_1$, according to the database state at time $t_2$?” Although Neo4j’s vector indexing infrastructure is less optimized for raw throughput than specialized vector databases, its property graph model and Cypher query language make bitemporal representations significantly easier to express and query.

\subsection{Graph-Based Retrieval}
GraphRAG~\cite{graphrag} and related work suggest that traversing knowledge graphs can improve retrieval by exposing relationships that aren’t explicit in the original query. In our case, the graph layer is more limited in scope: we maintain \texttt{RELATED\_TO} edges between memory identity nodes whenever a new embedding is sufficiently similar to an existing one (cosine similarity $\geq 0.75$). This effectively gives the agent a \texttt{get\_related\_memories} operation, where following an edge is often cheaper and more stable than reformulating the query and re-ranking results.

\subsection{Temporal Databases}
Bitemporal database theory distinguishes between valid time (when a fact is true in the real world) and transaction time (when it is recorded in the database)~\cite{snodgrass}. SQL:2011 introduced support for this model through period predicates and \texttt{FOR SYSTEM\_TIME AS OF} queries, but these features remain underused in most application-level databases. To our knowledge, a full two-axis temporal model has not yet been applied to a vector-indexed document store for LLM agent memory systems.

\section{System Design}

\subsection{Data Model}
\begin{figure}[ht]
\centering
\begin{tikzpicture}[
  node distance=1.1cm and 2.2cm,
  mem/.style={draw, rounded corners, fill=blue!8,
              minimum width=2.0cm, minimum height=0.6cm, font=\small},
  ver/.style={draw, rounded corners, fill=green!10,
              minimum width=2.6cm, minimum height=0.72cm, font=\small, align=center},
  old/.style={draw, rounded corners, fill=gray!12,
              minimum width=2.6cm, minimum height=0.72cm, font=\small, align=center},
  arr/.style={-Stealth, thick},
  lbl/.style={font=\scriptsize, midway}
]
\node[mem] (m) {\texttt{:Memory}};

\node[ver, right=2.2cm of m, yshift=0.7cm] (vcur)
  {\texttt{:MemoryVersion}\\[-1pt]{\tiny :CurrentVersion\ (tx\_to=null)}};

\node[old, right=2.2cm of m, yshift=-0.7cm] (vold)
  {\texttt{:MemoryVersion}\\[-1pt]{\tiny (closed)\ (tx\_to=$t_1$)}};

\draw[arr] (m.north east) -- node[lbl, above]{HAS\_VERSION} (vcur.west);
\draw[arr] (m.south east) -- node[lbl, below]{HAS\_VERSION} (vold.west);
\end{tikzpicture}
\caption{One \texttt{Memory} identity node with two content versions. The live version carries \texttt{:CurrentVersion}; the closed version retains its content and embedding permanently for time-travel queries.}
\label{fig:schema}
\end{figure}
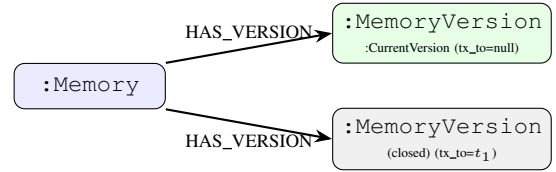

The schema separates \emph{identity} from \emph{content} so that memories can change over time without rewriting the graph structure (Fig.~\ref{fig:schema}). Concretely, each memory is stored as a stable \texttt{:Memory} node that only represents the memory’s identity, while the actual data is stored in separate \texttt{:MemoryVersion} nodes. Each time a memory is updated, a new version node is created and linked to the same identity node, rather than overwriting the previous one. This allows the system to preserve full history while keeping relationships fixed on the identity level.

Two node types are used:
\begin{itemize}
  \item \texttt{:Memory \{id\}} — this is a persistent identity node. It is never updated or deleted and serves as the anchor for both \textsc{RELATED\_TO} edges and the \textsc{HAS\_VERSION} chain.
  \item \texttt{:MemoryVersion \{...\}} — this node stores the actual content, including the embedding, category, tags, and four temporal timestamps. When the version is the current one, it also carries the \texttt{:CurrentVersion} label.
\end{itemize}

The schema uses two relationship types:
\begin{itemize}
  \item \texttt{HAS\_VERSION} — links a memory node to each of its content versions.
  \item \texttt{RELATED\_TO} — stores semantic similarity between memories and is created automatically at write time.
\end{itemize}

\subsection{Bitemporal Model}

Each \texttt{MemoryVersion} node stores two notions of time: \emph{valid time} and \emph{transaction time}. Valid time represents when a fact was true in the real world, while transaction time records when that fact was stored or modified in the database. Together, these timestamps allow the system to distinguish between when something happened and when the system learned about it.

Each dimension is represented as a closed-open interval:
\begin{align*}
\text{Valid time:}       &\quad [\,\texttt{valid\_from},\; \texttt{valid\_to}\,) \\
\text{Transaction time:} &\quad [\,\texttt{tx\_from},\;    \texttt{tx\_to}\,)
\end{align*}

A \texttt{NULL} upper bound indicates that the interval is still open. Valid time is provided by the caller and typically corresponds to when the memory was observed or extracted (e.g., the timestamp of the conversation session). Transaction time is assigned internally by the database: \texttt{tx\_from} records when the version was written, and \texttt{tx\_to} is set when that version is later updated or deleted.

On \texttt{update\_memory}, the current version is closed by setting its \texttt{tx\_to} timestamp and removing the \texttt{:CurrentVersion} label. A new version is then created with \texttt{tx\_from = now}. On \texttt{delete\_memory}, both \texttt{tx\_to} and \texttt{valid\_to} are closed on the active version. No versions are physically removed, which preserves the full history of the memory for temporal queries.

\subsection{Dual-Index Retrieval}

The system uses two separate HNSW vector indexes depending on the type of search being performed.

\paragraph{Current-state retrieval} The index \texttt{current\_version\_embedding} only contains nodes labeled \texttt{:CurrentVersion}. When a memory is updated, the old version loses this label and the new version receives it. As a result, searches on this index only return the latest version of each memory.

\paragraph{Historical retrieval}
The second index, \texttt{memory\_version\_embedding}, includes all versions of each memory, including older ones. To bound scan cost, the system over-fetches $10 \times k$ candidates from the vector index, then post-filters them using the valid-time and transaction-time conditions to keep only the versions that were active at the requested time with a query like below:

\begin{lstlisting}[language=SQL]
WHERE ($valid_at IS NULL OR (
    v.valid_from <= $valid_at AND
   (v.valid_to IS NULL OR v.valid_to > $valid_at)))
  AND ($tx_at IS NULL OR (
    v.tx_from <= $tx_at AND
   (v.tx_to IS NULL OR v.tx_to > $tx_at)))
\end{lstlisting}

Composite B-tree indexes on \texttt{(valid\_from, valid\_to)} and \texttt{(tx\_from, tx\_to)} accelerate this filtering step. The over-fetch factor of 10 is a deliberate tradeoff whose consequences are discussed in §V.

\section{Implementation}

\subsection{Technology Stack}

The backend uses Neo4j 5.27 Aura Enterprise for the main deployment, with local development targeting Neo4j Community via the default \texttt{bolt://localhost:7687} endpoint. The Python layer connects to the database using Neo4j’s official \texttt{neo4j} driver over the Bolt protocol.

For embeddings, we use Amazon Titan Embed Text v2 (\texttt{amazon.titan-embed-text-v2:0}) through AWS Bedrock, which produces 1024-dimensional, unit-normalized vectors. The agent is powered by Anthropic's Claude, and can be done either through the direct API or via AWS Bedrock. The model source can be switched at runtime using an environment variable.

\subsection{Agent Tool-Use Loop}

The agent uses Claude’s built-in tool-use loop, where the model decides which tools to call during a conversation. At each step, it outputs a \texttt{tool\_use} request, the system executes the requested operation on the memory store, and then returns the result as a \texttt{tool\_result}. The model then continues reasoning with this new information.

The system exposes nine memory-related tools, grouped below by their purpose:

\begin{itemize}
  \item \textit{Write operations:} \texttt{save\_memory}, \texttt{update\_memory}, \texttt{delete\_memory}
  \item \textit{Current-state retrieval:} \texttt{get\_memories}, \texttt{search\_memories}, \texttt{semantic\_search\_memories}
  \item \textit{Time-aware retrieval:} \texttt{as\_of\_semantic\_search}
  \item \textit{Graph-based access:} \texttt{get\_related\_memories}, \texttt{get\_memory\_history}
\end{itemize}

The \texttt{as\_of\_semantic\_search} tool adds a simple time filter using a \texttt{valid\_at} timestamp. This lets the agent ask questions about past states of memory. For example, a query like “what did I tell you about my diet last spring?” is translated into a specific date range, and the system returns only memory versions whose valid-time interval includes that date.

\subsection{Storing User Messages Only}

The system indexes only user messages and ignores assistant responses (i.e., any turn where \texttt{role != "user"}). This is a design choice based on the assumption that assistant outputs can be regenerated, while user inputs are the original source of information.

This design works well for user-centered memory retrieval, but it also means the system cannot answer questions about what the assistant previously said, since those responses are not stored. This limitation shows up in evaluation tasks that require recalling assistant-generated content.

\subsection{Automatic Edge Construction}

Each time a \texttt{save\_memory} or \texttt{update\_memory} call is made, we run a small follow-up query against \texttt{current\_version\_embedding} to retrieve the top-5 most similar memories using cosine similarity (with a cutoff of $\geq 0.75$). We then add or update \texttt{RELATED\_TO} edges between the corresponding identity nodes using those similarity scores.

This adds one extra ANN lookup per write, but in practice this overhead is small because writes happen much less often than reads in the agent loop. We chose a threshold of 0.75 for cosine similarity as lower values (around 0.7 or below) tended to connect memories that were only loosely related and made the graph noisier when we were testing.

\subsection{Legacy Data Migration}

On startup, \texttt{\_ensure\_schema} verifies that the required indexes exist. It removes any legacy flat index and creates the B-tree and HNSW indexes used by our current system. It also runs \texttt{\_migrate\_legacy\_memories} to update older stored memories.

Older entries stored memory content directly on the \texttt{:Memory} node without embeddings. During migration, these entries are converted to the current structure by creating a \texttt{:MemoryVersion:CurrentVersion} node, generating an embedding from the original text, and removing the raw content from the identity node.

This migration runs once during initialization and has no effect on subsequent runs.

\section{Evaluation}

\subsection{Benchmark and Protocol}

We evaluated our implementation on LongMemEval~\cite{longmemeval}, a 500-question benchmark built
from synthetic multi-session conversations. The questions on this benchmark fall into six types:
single-session user statements (\textit{ss-user}), assistant outputs
(\textit{ss-asst}), inferred preferences (\textit{ss-pref}), facts spread
across sessions (\textit{multi-session}), date arithmetic (\textit{temporal-reasoning}),
and facts that changed over time (\textit{knowledge-update}).

For each example, we clear the database and ingest all user turns with at least 20 characters, setting \texttt{valid\_from} to the session date. We then retrieve answers using two retrieval modes. The default mode uses vector similarity search with the top-10 results (Strategy 2). For \textit{temporal-reasoning} and \textit{knowledge-update} questions, we additionally use a time-aware retrieval mode that filters results by \texttt{valid\_at} (Strategy 1). A hit requires
$\geq$50\% token overlap with the ground-truth answer. We sample 10 examples
per type (60 total, seed=42).

Results are reported using R@$k$ (recall at $k$), which measures whether the correct answer appears within the top $k$ retrieved results.

\subsection{Results}

\begin{table}[ht]
\centering
\caption{LongMemEval results. R@$k$ = Strategy~2 (current-state).
R@10$_t$ = Strategy~1 (time-travel), reported only for the two types where it runs. Strategy~1 returned zero candidates for some temporal-reasoning questions; those are excluded from the R@10$_t$ denominator (see §V-E).}
\label{tab:results}
\setlength{\tabcolsep}{4.5pt}
\begin{tabular}{@{}lccccc@{}}
\toprule
Question Type        & N  & R@1  & R@5  & R@10 & R@10$_t$ \\
\midrule
single-session-user  & 10 & 70.0 & 90.0 & 90.0 & ---  \\
knowledge-update     & 10 & 40.0 & 80.0 & 80.0 & 80.0 \\
temporal-reasoning   & 10 & 50.0 & 50.0 & 50.0 & 37.5$^\dagger$ \\
multi-session        & 10 & 0.0  & 30.0 & 30.0 & ---  \\
single-session-asst  & 10 & 0.0  & 20.0 & 20.0 & ---  \\
single-session-pref  & 10 & 0.0  & 10.0 & 10.0 & ---  \\
\midrule
\textbf{Overall}     & 60 & 26.7 & 46.7 & 46.7 & ---  \\
\bottomrule
\end{tabular}

\vspace{2pt}
{\footnotesize $^\dagger$ Computed over 8 non-null results; 2 correctly returned empty (§V-E).}
\end{table}

\subsection{Single-Session User Statements}

At 90\% R@10 this is the strongest category. Questions ask about things the
user said directly, and the corpus \emph{is} the user's own words, so semantic
search finds them reliably. Notably, R@5 equals R@10 for every question type,
meaning every hit found in the top 10 was already present in the top 5—the
1024-dimensional Titan embeddings consistently rank correct matches highly
when they exist in the corpus.

\subsection{Single-Session Assistant and Preference Types}

Both low scores reflect indexing choices, not retrieval quality.
\textit{ss-asst} (20\%) fails because we only index user turns (§IV-C), i.e.
the assistant's recommendations are never stored.
\textit{ss-pref} (10\%) performs poorly because the correct answers are synthetic multi-sentence summaries that do not appear directly in any user message, so they cannot be retrieved exactly through search alone.

\subsection{Temporal Reasoning and the Dilution Effect}

Current-state search achieves 50\% R@10, while time-travel search achieves 37.5\%. Two of the ten examples produced null for R@10$_t$ rather than false: Strategy~1 returned zero results because the relevant session's date fell after the question's \texttt{valid\_at}. This is correct temporal behavior—the memory did not yet exist at the point in time the question specifies—and it is treated as a correct empty result, not a miss. Those two examples are excluded from the 37.5\% denominator.

For the remaining eight examples, the current-state path scored $5/8 = 62.5\%$ while time-travel scored $3/8 = 37.5\%$. The time-travel path performs worse despite having access to the same or more data. The cause is the over-fetch strategy: to allow post-filtering, we pull $10 \times k = 100$ candidates from the full-history HNSW index. Many of those candidates are older versions of memories that pass ANN scoring but fail the temporal filter, and the survivors are re-ranked by similarity without any recency signal. The result is that the target answer sometimes drops below position 10 in the filtered set, even though it would have been in the top 5 under the current-state path. This is a concrete, reproducible failure mode of the over-fetch design rather than a problem with the temporal model itself.

\subsection{Knowledge Update}

Both strategies achieve 80\% R@10. In these cases, the updated values were still valid at the question time, so time-travel retrieval returned the same results as current-state search. The two missed cases required combining values across multiple sessions rather than retrieving a single updated fact.

\subsection{Multi-Session}

At 30\% R@10 and 0\% R@1, multi-session is the weakest non-structural category. Most ground-truth answers require counting events across multiple sessions (e.g., “how many farmers market trips did you take?”). These cannot be answered by retrieval alone. The 30\% that succeed are cases where the answer is explicitly stated in a single user message.

\section{Conclusion}

\subsection{Summary}

In this project we implemented a conversational memory store on Neo4j that combines vector-based similarity search with a bitemporal data model and automatic graph linking between related memories. The main components of the system are: (1) separating identity nodes from versioned memory data so that temporal queries can be expressed directly in Cypher while still supporting HNSW search; (2) using Neo4j labels to support both a fast “current state” view and a full history view over the same data; and (3) adding composite B-tree indexes over the bitemporal interval columns to speed up filtered queries.

We evaluated the system using LongMemEval. It performs well on direct factual recall (about 90\% on user-statement questions and 80\% on knowledge-updates). It is less reliable on tasks that require combining information across multiple sessions, inferring user preferences, or recalling indirect assistant-generated content. We also found that time-based retrieval can pull in too much unrelated context, which hurts performance on temporal reasoning tasks. This highlights a trade-off between retrieving more information and keeping results focused, and it is an actionable finding rather than a fundamental limit of the temporal model.

\subsection{Future Directions}

\textbf{Re-ranking after filtering.} To reduce noise in temporal reasoning, we can re-rank retrieved memories after filtering by combining cosine similarity with closeness to \texttt{valid\_at}, instead of relying only on the initial vector search score.

\textbf{Event aggregation during ingestion.} For questions that involve counting across sessions, we likely need to process events at ingestion time by extracting entities and updating counter nodes in the graph. This would store totals directly instead of relying only on raw text.

\textbf{Indexing assistant messages.} Supporting recall of assistant-generated content requires also indexing assistant turns during ingestion. These can be tagged separately from user messages and handled differently at retrieval time using the existing \texttt{category} field.

\end{document}